\documentclass[twocolumn,prl,showpacs]{revtex4}

\usepackage{epsfig}

\begin{document}

\title{Observation of electronic and atomic shell effects in gold nanowires}

\preprint{7}

\author{A.\,I.\  Mares}
\affiliation{Kamerlingh Onnes Laboratorium, Universiteit Leiden,
Postbus 9504, NL-2300 RA Leiden, The Netherlands}

\author{A.\,F.\  Otte}
\affiliation{Kamerlingh Onnes Laboratorium, Universiteit Leiden,
Postbus 9504, NL-2300 RA Leiden, The Netherlands}

\author{L.\,G.\ Soukiassian}
\affiliation{Kamerlingh Onnes Laboratorium, Universiteit Leiden,
Postbus 9504, NL-2300 RA Leiden, The
Netherlands}

\author{R.\,H.\,M.\  Smit} \affiliation{Kamerlingh
Onnes Laboratorium, Universiteit Leiden, Postbus 9504, NL-2300 RA
Leiden, The Netherlands}

\author{J.\,M.\ van Ruitenbeek}
\affiliation{Kamerlingh Onnes Laboratorium, Universiteit Leiden,
Postbus 9504, NL-2300 RA Leiden, The Netherlands}

\date{\today}

\begin{abstract}
The formation of gold nanowires in vacuum at room temperature
reveals a periodic spectrum of exceptionally stable diameters.
This is identified as shell structure similar to that which was
recently discovered for alkali metals at low temperatures. The
gold nanowires present two competing `magic' series of stable
diameters, one governed by electronic structure and the other by
the atomic packing.

\end{abstract}


\maketitle

When the conductance of metallic nanowires can be described in
terms of a finite number of quantum modes it is expected that the
degree of filling of the quantum modes has a measurable effect on
the total free energy of the nanowires
\cite{ruitenbeek97,stafford97,yannouleas97}. Indeed, Yanson {\it
et al.} \cite{yanson99} observed that alkali nanowires, obtained
by the mechanically controllable break junction (MCBJ) technique,
exhibit exceptional stability for certain diameters. The series of
stable diameters was shown to correspond directly to the
well-known magic number series of alkali metal clusters produced
in vapor jets (for a review see \cite{deheer93}). For alkali metal
clusters and nanowires the main features of this electronic shell
effect can be understood in terms of the filling of simple
free-electron modes inside a spherical or cylindrical cavity,
respectively. The electronic level spectrum shows groups of levels
bunched together and a cluster or nanowire is favored when such a
shell is just filled. The principle of this shell structure is
well known from atomic physics, giving rise to the periodic table
of the elements, and from nuclear physics, explaining the periodic
appearance of stable nuclei. Apart from this electronic shell
structure alkali metal clusters and nanowires show, typically at
larger diameters, a second series of stable diameters resulting
from the atomic structure. Completed atomic layers stabilize the
nanowires.

The alkali nanowires are very attractive from a fundamental point
of view, since the electronic structure is well-described in terms
of free and independent electrons, but they are not very suitable
for possible applications. As suggested by research on metal
clusters shell effects for nanowires should be observable for many
other metals. As a first step beyond the alkali metals we have
chosen gold, since the physics is expected to be very similar, it
being to a good approximation still a free-electron metal.

Recently some experimental results on the formation and stability
of gold nanowires have already been reported. By the use of an
electron beam thinning technique in a transmission electron
microscope under ultra-high vacuum at $T=300$\,K Kondo and
Takayanagi \cite{kondo00} observed gold nanowires that adopt a
series of unusual helical configurations, which is an exceptional
form of atomic shell structure. Using STM indentation experiments
Medina {\it et al.} \cite{medina03,diaz03} reported the
observation of atomic shell effects in gold nanowires. Here, we
demonstrate clear evidence for {\em electronic} shell effects,
with a cross-over to atomic shell structure at larger diameters.

\begin{figure}[t!]
\begin{center}
\epsfig{width=7cm,figure=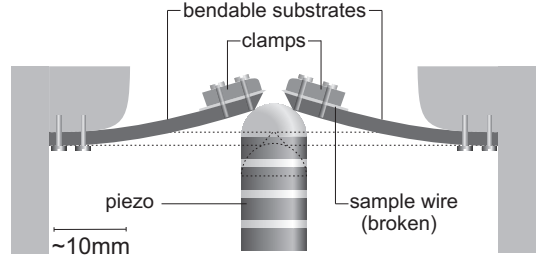}
\end{center}
\vskip -0.5cm \noindent{\caption{Schematic view of the MCBJ
technique under UHV. The sample wire is clamped onto two separate
bending beams, having a notch at the bridging point, and broken by
bending the beams. Contact between the fracture surfaces can be
finely adjusted by means of the piezoelectric element. The relaxed
configuration of the sample is represented by the dotted lines.
 } \label{bendratioPRL}}
\end{figure}

The atoms need to have sufficient mobility in order to be able to
explore many wire configurations and find the stable ones
corresponding to shell filling. Therefore, in order to observe
shell effects we need to raise the temperature to a sizeable
fraction of the bulk melting temperature. For the alkali metals a
temperature near 80\,K was sufficient and the experiments were
performed with a standard low-temperature MCBJ device with a local
heating extension \cite{yanson99}. For gold this is not sufficient
and we have developed a new MCBJ instrument
(Fig.\,\ref{bendratioPRL}) operating under ultra high vacuum (UHV)
allowing the temperature to be varied between 70\,K and 500\,K. A
notched gold wire is fixed with stainless steel clamps, bridging
two bendable metal substrates. The wire is broken at the notch in
a controlled way by mechanically applying pressure from below onto
both substrates. Once broken the bending beams can be relaxed to
bring the wire ends back into contact and atomic-sized contacts
can be finely adjusted using a piezoelectric element. The base
pressure is $5 \times 10^{-10}$\,mbar.

Nanowires are obtained by indenting the two clean fracture
surfaces and subsequently stretching the contact while monitoring
the conductance of the wire. The conductance trace follows a
sequence of plateaus separated by abrupt jumps. At the final
stages the contact consists of just a few atoms and the
conductance is changing in steps signaling the atomic structural
rearrangements of the contact. For monovalent metals the
conductance of a single-atom contact is close to
1\,$G_{0}=2e^{2}/h$, and carried by a single conductance channel
\cite{scheer98}. The conductance is measured at constant bias
voltage, recording the current through a current-voltage convertor
and a 16 bit analog-to-digital convertor. A trace from 100\,$G_0$
to 0 is recorded in about 10\,s at a sampling rate of 10\,ms.

At larger diameters the conductance jumps can be larger and each
conductance trace is different depending on the evolution of
atomic arrangements in the nanowire. In order to reveal
underlaying variations in stability we perform a simple
statistical analysis by adding all values of the digitized
conductance traces into conductance histograms. Preferred nanowire
structures are then seen as maxima in the conductance histograms.
The wire radius $R$ corresponding to a given conductance value can
be obtained from a semi-classical expression
\cite{torres94,hoppler98}:
\begin{equation}\label{Sharvin}
  G\cong G_{0}\left[\left(\frac{k_{\rm F}R}{2}\right)^{2}-\frac{k_{\rm F}R}{2}+\frac{1}{6}\right],
\end{equation}
with $k_{\rm F}$ the Fermi wave vector.

\begin{figure}[t!]
\begin{center}
\epsfig{width=8cm,figure=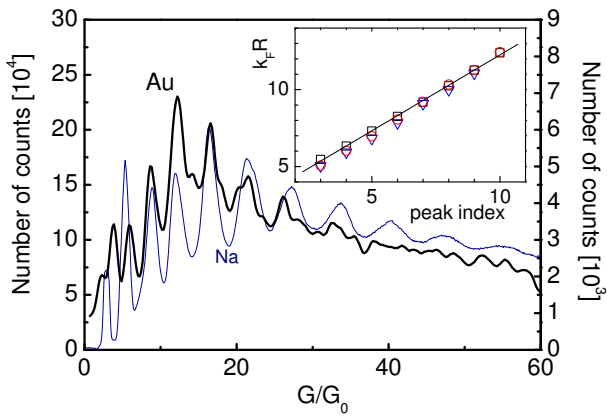}
\end{center}
\vskip -0.7cm \caption{(color online) Conductance histogram for
gold (bold curve, right axis) giving evidence for electronic shell
structure. It is obtained at room temperature from 2000 individual
consecutive traces, using a bin-size of $\sim$0.1\,$G_0$ and a
bias voltage of 300\,mV. For comparison similar data from
\protect\cite{yanson99} obtained for Na at 80\,K are shown (thin
curve, left axis). The inset shows the peak positions, converted
to $k_{\rm F}R$, as a function of peak index for three independent
experiments on Au.} \label{histESE}
\end{figure}

Figure\,\ref{histESE} shows a conductance histogram up to a
conductance of 60\,$G_0$ obtained for gold at room temperature
(bold curve). The histogram is five-point smoothed by using a
second order polynomial. The data show strong resemblance to the
histogram for Na recorded at $T=80$\,K by Yanson {\it et al.}
\cite{yanson99}, which is shown for comparison in the same graph.
The high counts at the peaks result from stable wire diameters.
The peaks between 7 and 25\,$G_0$ agree very well. Above 25\,$G_0$
only one or two more peaks are visible for Au, much less than for
Na, and they may be shifted somewhat to lower conductance. The
first few peaks below 7\,$G_0$ do not correspond very well.  The
low-conductance histograms for Na have been shown before to differ
significantly from those for Au \cite{brandbyge95,krans95}. The
direct correspondence in period and phase of the two series of
peaks at higher $G$ is strong evidence for similar electronic
shell structure in Au and Na.

For shell effects the peaks are expected to be equidistant when
plotted as a function of the radius of the wire. This is verified
in the inset to Fig.~\ref{histESE}, where we plot $k_{\rm F}R$
obtained from Eq.\,(\ref{Sharvin}) as a function of peak index.
The peaks have been reproduced for different samples in six
independent series of measurements on Au and in order to
illustrate the degree of reproducibility the plot contains points
from three measurements. The period $\Delta k_{\rm F}R$ shows some
variation between the individual measurements that averages as
$\Delta k_{\rm F}R=1.02\pm0.04$. This is somewhat smaller than the
slopes for electronic shell structure observed for the alkali
metals \cite{yanson01}.

\begin{figure}[t!]
\begin{center}
\epsfig{width=8cm,figure=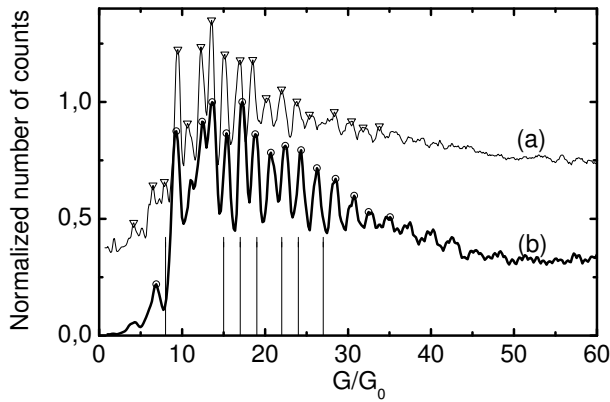}
\end{center}
\vskip -0.7cm \noindent{\caption{Second type of conductance
histogram for Au at room temperature, giving evidence of
electronic and atomic shell effects with a crossover at
G$\simeq$12\,G$_0$. (a) Obtained from 3000 individual conductance
traces recorded at a bias voltage of 150\,mV. (b) Recorded at
100\,mV, combining 5000 traces, evidencing mainly atomic shell
effect, since only the first two peaks can be related to
electronic shell effect. For comparison the positions of the
stable helical gold nanowires proposed by Kondo {\it et al.}
\protect\cite{kondo00} are marked by vertical
lines.}\label{ASEfinal}}
\end{figure}

\begin{figure}[t!]
\begin{center}
\epsfig{width=7cm, figure=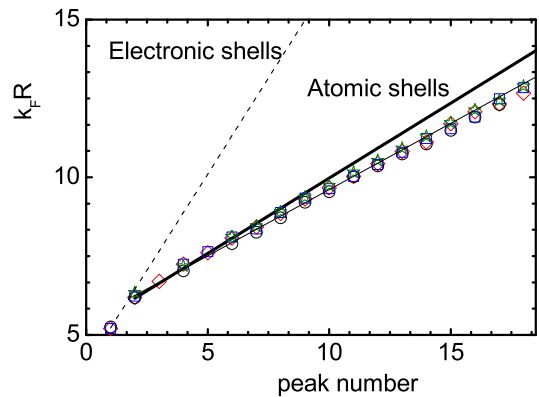}
\end{center}
\vskip -0.5cm \noindent{\caption{(Color online) Positions of the
peaks in the histograms for four independent experiments,
including those from Fig.~\protect\ref{ASEfinal}, reproducibly
showing evidence for a second set of periodic stable diameters.
The dashed line gives the expected slope for electronic shell
structure; the bold line gives the slope for the stable radii of
atomic shells expected from close packing of the nanowire along
the $[110]$ axis, $\Delta k_{\rm F}R=0.47$. In the experiment the
slope above the second point is $\Delta k_{\rm F}R=0.4\pm0.01$.
This is somewhat smaller (see text) but suggests an explanation in
terms of atomic shell structure. } \label{linebest}}
\end{figure}

Often we observe another periodic structure in gold conductance
histograms recorded under similar experimental conditions, as
illustrated by two examples in the Fig.~\ref{ASEfinal}.  Many
histograms show a cross-over between the electronic shell period
of Fig.\,\ref{histESE} to a new shorter period. The peaks were
identified using a software tool (symbols) and Fig.~\ref{linebest}
shows the radii corresponding to the peaks in
Fig.~\ref{ASEfinal}(a). The cross-over is at the second point,
above which we find a period $\Delta k_{\rm F}R=0.40\pm 0.01$.
This structure has been reproduced in well over ten independent
measurements and a second example is shown in
Fig.~\ref{ASEfinal}(b) for comparison. It can be seen that the
period $\Delta k_{\rm F}R$ displays some variation, giving an
overall average of $\Delta k_{\rm F}R=0.40\pm0.03$. Again, this
second periodic structure agrees closely to the results obtained
for alkali nanowires and by analogy we interpret it as an {\em
atomic} shell effect.

Thus we attribute the peaks in the histograms in
Figs.\,\ref{histESE} and \ref{ASEfinal} to two different and
independent effects: electronic and atomic shell effects. The
electronic shell effect gives rise to peaks in the histograms
because specific radii of the nanowire are more stable than
average due to local minima of the electronic energy of the
system. Electronic shell effects have been previously found for
clusters of gold atoms in vapor jets \cite{katakuse85}.  The magic
number series of stable clusters has been successfully explained
by approximating the electronic structure of gold metal by
homogeneous free-electron (jellium) models \cite{deheer93,brack93}
confined by a spherically symmetric potential well. In the same
spirit we model our nanowires as a free electron gas confined by a
cylindrical potential well
\cite{stafford97,yannouleas97,hoppler98,ogando02}. The magic
diameters are obtained by a semiclassical approach using the
Bohr-Sommerfeld quantization condition. Stable trajectories for
the electron wavepackets are periodic orbits inside a cylinder,
that can be classified as diametric, triangular, square, etc. Each
of these trajectories gives rise to a periodic series of
oscillations in the density of states as a function of $k_{\rm
F}R$, that produce oscillations in the total energy of the system.
The dominant period of the oscillation depends somewhat on the
electron density and the shape of the potential well. We obtain a
number of $\Delta k_{\rm F}R=1.21\pm 0.02$ using an electron
density corresponding to that for gold from the calculations by
Ogando {\it et al.} \cite{ogando02}, which agrees well with the
periodicity of shell structure in gold clusters, $\Delta k_{\rm
F}R=1.23\pm 0.01$. The differences in the peaks observed for Na
and Au at low $G$ can be largely explained by considering
elliptically distorted wires and including differences in surface
tension for the two metals \cite{urbanTBP}. However, we find a
reproducible peak at 4\,$G_0$ for gold, which does not appear in
this analysis and may be due to quadrupolar distortions
(\cite{urbanTBP} and private communications).

The period obtained experimentally for gold nanowire magic radii,
$\Delta k_{\rm F}R=1.02\pm 0.04$, is slightly smaller than that
expected from the semiclassical analysis. Deviations similar to
the present results for Au have been observed for Li
\cite{yanson01}. Such deviations have been attributed to
backscattering of the electrons on defects, which is not included
in the idealized semiclassical model. Defect scattering may also
explain the observed variation of the slopes between experiments
and the fact that the range of observed oscillations for Au is
smaller than for Na, since scattering will lead to smearing of the
peaks. In order to preserve a constant slope the correction should
be proportional to the conductance $G$ itself. In other words, a
certain fraction of the electrons is scattered back, independent
of the wire radius. This would for instance be the case when the
dominant mechanism is scattering by roughness on the nanowire
surface, which seems plausible.

The Fermi surface of bulk gold has marked deviations from
spherical symmetry. The nearly spherical main sheets of Fermi
surface are connected by necks across the Brillouin zone
boundaries. The necks may introduce an additional set of
oscillations in the density of states. However, since the
wavelength of the states in the neck is about six times larger
than the main Fermi wavelength the period would be much larger
than that due to the main states. Moreover, the number of states
in the neck is small making the amplitude of the effect small.
Thus we expect that it is a good approximation to consider gold as
free electron metal having the same electronic shell effect as in
alkali metals. This is supported by electronic structure
calculations \cite{opitz02} for the quantum modes in nanowires of
Na and Cu (which has a Fermi surface similar to Au).

For larger radius $R$ the configuration energy (atomic shell
structure) becomes more important than the energy contribution due
to the conduction electrons. The amplitude of the oscillations in
the electronic free energy scale as $1/R$ \cite{yannouleas98},
while the ones for the surface energy increase proportional to
$R$, making the latter dominant at large radii. The cross-over
point is seen to vary between experiments, which is likely due to
differences in the local crystal orientation of the leads
connecting the nanowire. Note that the histogram in
Fig.\,\ref{histESE} shows some additional fine structure that may
be due to an admixture of atomic shell structure.

The [110] orientation has been shown to be most favorable to form
nicely faceted long nanowires \cite{rodrigues00,jagla01}. At large
diameters we expect the nanowires to order into densely packed
wires, that minimize the surface energy. We assume that the bulk
fcc packing of gold is preserved, which is supported by the
observation of gold nanowires in transmission electron microscopy
by Rodrigues {\it et al.} \cite{rodrigues00}. The lowest energy
surfaces for gold are perpendicular to $[111]$ and we can
construct a densely packed wire along a $[110]$ axis with four
$(111)$  and two $(100)$ facets, as also proposed by
Ref.\,\onlinecite{jagla01}. Completing a full atomic shell gives
rise to a rather long period of $\Delta k_{\rm F}R=2.85$, but
assuming stable wires are found when completing a single facet we
obtain a six times smaller period $\Delta k_{\rm F}R=0.476$
\cite{yanson01a}. The faceting structure of stable wires can be
recognized in a Monte Carlo simulation of the thinning-down of a
fcc wire by Jagla and Tosatti \cite{jagla01}. The experimental
slope is about 20\% smaller than expected, but this is consistent
with a similar deviation observed for the electronic shell effect
and is similarly attributed to defect scattering.

Our results and interpretation differ in several essential points
from the work in Ref.\,\onlinecite{diaz03}, where the observed
structure is entirely attributed to atomic shell structure.
However, the well-developed peak numbers 3, 4 and 5 in Fig.\,1 of
Ref.\,\onlinecite{diaz03} agree with the first three peaks of what
we interpret as electronic shell structure in Fig.\,\ref{histESE}.
We have no need to resort to a second-derivative analysis, as was
done in the cited work to analyze the histograms at higher $G$,
which entails the risk of identifying residual fluctuations in the
background as peaks. Also, our peak structure shows the expected
global decrease of the amplitude towards larger diameters, in
contrast to the data by Medina {\it et al.}
\cite{medina03,diaz03}. We conclude that electronic shell
structure is clearly observed in Au nanowires, with a cross over
to atomic shells at larger diameters.

Apart from atomic shell structure based on conventional fcc
lattice packing one should consider the `weird wire'
configurations that have been predicted \cite{gulseren98} and
observed for gold in transmission electron microscopy
\cite{kondo00}. The helical atomic wire arrangements lead to a
series of `shells' for which we estimate the conductances shown by
bars in Fig.\,\ref{ASEfinal}. In view of the much more regular
pattern observed we conclude that under conditions used in our
experiment the regular lattice atomic shell structure is
predominant. This may be a consequence of the path by which the
nanowire evolves from a large fcc wire down. Making a structural
cross-over to helical arrangement along the way presumably
involves a relatively large barrier.

We thank R. van Egmond for valuable assistance with the
experiments, and J. B{\"u}rki, H. Grabert, and D. Urban. This work
is part of the research program of the ``Stichting FOM,''and was
supported by the EU TMR Network program DIENOW.

\bibliographystyle{apsrev}
\bibliography{D:/user/papers/bib/QPC_v26}

\end{document}